\documentclass[a4paper,11pt]{article}
\usepackage[german,american]{babel}
\usepackage[utf8]{inputenc}
\usepackage[a4paper,top=1.8cm,bottom=2.0cm,left=3cm,right=3cm,marginparwidth=1.75cm]{geometry}
\usepackage{amsmath}
\usepackage{amsfonts}
\usepackage[T1]{fontenc}
\usepackage{amssymb}
\usepackage{graphicx}
\usepackage{caption}
\usepackage{subcaption}
\usepackage{float}
\usepackage{amsmath} 
\usepackage[OMLmathsfit]{isomath}
\usepackage{amssymb} 
\usepackage{amsxtra} 
\usepackage{bm} 
\usepackage{mathtools}
\usepackage[usenames]{color}
\usepackage{verbatim}
\usepackage[colorlinks=true, allcolors=blue]{hyperref}
\usepackage[dvipsnames]{xcolor}
\usepackage{multirow}
\usepackage[section]{placeins}

\usepackage{booktabs} 
\usepackage{siunitx}  

\usepackage[numbers,sort&compress]{natbib}

\definecolor{darkgreen}{rgb}{0.0,0.5,0.0}
\definecolor{BurntOrange}{rgb}{0.8,0.3,0.0}
\definecolor{mygray}{gray}{0.5}

\def\msbi#1{\mathsfbfit{#1}}

\providecommand{\keywords}[1]
{
  \small	
  \textbf{\textit{Keywords---}} #1
}

\title{Lossless Multi-Scale Constitutive Elastic Relations \\ with Artificial Intelligence}

\author{Jaber Rezaei Mianroodi$^{1,*}$, Shahed Rezaei$^{2}$, Nima H. Siboni$^{3}$,\\ Bai-Xiang Xu$^{2}$,  Dierk Raabe$^{1}$ \\
       \footnotesize $^1$Microstructure Physics and Alloy Design, \\ 
        \footnotesize Max-Planck-Institut f\"ur Eisenforschung, D\"usseldorf, Germany \\
        \footnotesize $^2$Mechanics of Functional Materials Division, Institute of Materials Science,\\ \footnotesize Technische Universität Darmstadt, Darmstadt, Germany \\
        \footnotesize $^3$DeepMetis, Lohm\"uhlenstraße 65, 12435 Berlin, Germany \\
        \footnotesize $^*$Corresponding Author j.mianroodi@mpie.de
} 

\begin{document} 

\maketitle

\begin{abstract} 

The elastic properties of materials derive from their electronic and atomic nature. However, simulating bulk materials fully at these scales is not feasible,  so that typically homogenized continuum descriptions are used instead. A seamless and lossless transition of the constitutive description of the elastic response of materials between these two scales has been so far elusive. Here we show how this problem can be overcome by using Artificial Intelligence (AI). A Convolutional Neural Network (CNN) model is trained, by taking the structure image of a nanoporous material as input and the corresponding elasticity tensor, calculated from Molecular Statics (MS), as output. Trained with the atomistic data, the CNN model captures the size- and pore-dependency of the material's elastic properties which, on the physics side, can stem from surfaces and non-local effects. Such effects are often ignored in upscaling from atomistic to classical continuum theory.
To demonstrate the accuracy and the efficiency of the trained CNN model, a Finite Element Method (FEM) based result of an elastically deformed 
nanoporous beam equipped with the CNN as constitutive law is compared with that by a full atomistic simulation.  
The good agreement between the atomistic simulations and the FEM-AI combination for a system with size and surface effects establishes a new lossless scale bridging approach to such problems. 
The trained CNN model deviates from the atomistic result by 9.6\% for porosity scenarios of up to 90\% but it is about 230 times faster than the MS calculation and does not require to change simulation methods between different scales. 
The efficiency of the CNN evaluation together with the preservation of important atomistic effects makes the trained model an effective atomistically-informed constitutive model for macroscopic simulations of nanoporous materials, optimization of nano-structures, and solving of inverse problems. 

\end{abstract} 


\section{Introduction}

The laws of mechanics are rather well established at the macroscopic continuum scale.
This applies particularly for the linear relationship between stress and strain, such as cast in Hooke's linear elasticity law nearly 350 years ago.
The material-specific properties are represented through the elastic stiffness modulus,  which in three dimensions is a 4th order tensor. 
However, when leaving the idealized continuum description of bodies and zooming into their underlying micro-cosmos, also referred to as microstructure, a very complex landscape emerges which is characterized by a wide range of structural features and defects.
Some of these microstructure features do not alter the macroscopic stiffness substantially or only in a rather modest and linear fashion, but others lead to more drastic changes of the elastic response.
This applies particularly to the porosity of the material, a property which refers to the volume fraction, dispersion and connectivity of the open volume.
In such cases the elastic modulus of the representative volume element of a material can be altered quite substantially. The relationship between porosity and elastic stiffness as well as its size-dependence across several scales is of importance for a number of material classes. 
One example is bone, where diseases such as osteoporosis lead to loss in bone mass through the highly dispersed micro-architectural deterioration of bone tissue, an effect which entails decay in elastic stiffness and bone fragility \cite{Strom2011}. A direct relationship between porosity across all size scales and the resulting stiffness and strength also applies for wood, which is currently gaining rapid momentum as sustainable construction material \cite{Wimmers2017}. Another example is the development of nano- and micro-porosity during the direct reduction of iron ores \cite{KIM2021116933}. This effect leads to gradual loss in stiffness which has severe consequences for the design of corresponding direct reduction and fluidized bed reactors, a field of utmost relevance for the carbon free reduction of metal oxides.  

Also many functional materials have a nano- or micro-porous structure, such as aerogels, many catalysts or supercapacitors \cite{CHOWDHURY2019265}. Most of these materials do not only fulfill their respective functional role but must at the same time bear mechanical loads, where the elastic stiffness assumes a central role in their property portfolio. Another field where porosity and stiffness are closely connected is the domain of additive manufacturing which plays an increasing role in digital manufacturing. Most parts cannot be manufactured to 100\% mass density so that better understanding of corresponding stiffness effects due to the inherited porosity is an important aspect \cite{ZHANG2019497}.

A size-dependency of the elastic properties is particularly expected when it comes to nanoporous materials which are attributed to the competition between surface and bulk energies, especially at smaller scales \cite{LUCCHETTA2021103669, Wang2006, Chen2014}. Classical models of elasticity should be extended correspondingly through the incorporation of surface elasticity models \cite{Shuttleworth_1950, Gurtin, Javili2014}. The reader is referred to \cite{LI20061195, Firooz2021,Wei_2006, STEIN2016154}, and the references therein for an overview of surface effects in case of elasticity and plasticity.
One effective approach is to utilize interface finite elements on the surface of models that represent nanostructured materials. The latter method requires explicit input of the surface tension constants and surface elastic parameters in the case of surface elasticity. Also, such surface elements induce implementation and computational complications \cite{Firooz2021, Rezaei2020}. 
A size effect is also predicted in the torsion and bending of open-cell foams or lattices, where slender specimens appear stiffer than expected. Such size effect can be predicted by Cosserat elasticity (also known as micropolar elasticity). Such models are sensitive to strain gradients and introduce a characteristic length into the constitutive formulations which, however, require complex experimental characterization to identify the required constitutive parameters (see also \cite{Rueger2019}). When it comes to the mechanical and topological homogenization of such materials, further extensions are required to calibrate an effective generalized continuum model that is applicable across several scales \cite{Forest2011, Kouznetsova2002, Geers2003}. 

A straightforward way to calculate the true elastic stiffness of nanoporous materials is conducting atomistic calculations which naturally consider all relevant porosity and surface tension features arising form it \cite{GUILLOTTE2019135, Patil2017}.
However, this is not feasible when targeting the mechanical description of larger parts, revealing the typical scale-dilemma often encountered in computational materials science. This problem is also often referred to as the multi-scale and multi-physics challenge in materials mechanics \cite{van_der_Giessen_2020}. 

The multi-scale modeling strategy requires, on one hand, accurate and predictive simulation at lower scales taking various physical phenomena into account, and on the other hand, efficient methodologies to transfer the information between the scales. 
In the available scale-bridging techniques, there is usually a trade-off between the amount of information preserved in the up-scaling and the associated computational costs. One way to establish this micro-macro coupling is to make use of Artificial Intelligence (AI). Machine Learning (ML) seems to provide a promising approach for an efficient scale-bridging.
As mentioned, the properties of a material depend to a large extend on a wide cosmos of defects, which is also referred to as nano- and microstructure, and on the mutual interaction among all these features.
The high-dimensional and tensorial interaction among these multiple nano- and microstructure features makes the required dimensionality reduction of the scale transfer problem much more complex than just transferring the chemical composition and some overall geometrical factors from one scale to another. 
Empirical descriptors that could be predefined to capture and reflect some of these features and their change upon scale transitions are usually unknown and difficult to determine. Instead, in most cases, there are more complex and sometimes even weakly understood interactions hidden inside the material's nano- and microstructure. Therefore, image-based AI methods can be valuable tools for the study of microstructure-dependent structure-property relations \cite{mianroodi2021teaching, LIN2021109193} in general (due to the many effects and phenomena involved) and specifically for scale-bridging structure-property calculations (where even some of the physics for adequate coarse-graining have not been resolved yet). In \cite{Huang2020}, a summary is provided over recent advances in the application of AI techniques for numerical modeling of various types of materials such as metals, polymers, ceramics and composites. 
Owing to their high efficiency, allowing very fast calculations, AI techniques open up new efficient strategies to optimize and drastically accelerate structure-property calculations of future advanced engineering materials and structures \cite{SALEHI2018170}. These approaches could also help to discover scale bridging phenomena that had so far remained elusive, hidden behind the enormous chemical and structural complexity of modern engineering materials.
Peng et al. \cite{Peng2020} discussed the state of the art of combining ML and multi-scale modeling in various applications. Bock et al. \cite{Bock2019} reviewed successful applications of ML and statistical learning methodologies in the field of continuum materials mechanics. They concluded that simulation-based data mining in combination with ML tools provide exceptional opportunities for identification of fundamental interrelations within materials. 

Different ML strategies have been investigated for modeling at different length scales. 
Wang and Sun \cite{WANG2018337} replaced the up-scaling procedure through an offline homogenization procedure by utilizing a sub-scale simulations to generate a database to train material models for geological materials. Xue et al. \cite{D0SM00488J} proposed a data-driven multi-scale computational scheme to capture the non-linear mechanical behavior of cellular meta-materials. See also \cite{Yvonnet2015} for similar studies.
Kumar et al. \cite{Kumar2020} introduced a ML technique for the inverse design of meta-materials which is able to generate functionally graded cellular structures with tailored anisotropic stiffness.
Wang et al. \cite{Wang2020} developed a data-driven method for an efficient multi-scale topology optimization. 
The application of ML has also been extended to more complicated material behavior including plasticity
\cite{SETTGAST2020102624, FUCHS2021106505}. 
Readers are referred to \cite{KARAPIPERIS2021104239, LIN2021109193} for similar studies in different application fields. 
It is worth to mention that the ML approaches can be applied to construct a direct solver for the usually well-known Partial Differential Equations (PDEs) out of massive data sets.
Samaniego et al. \cite{SAMANIEGO2020112790} explored deep neural networks as an option to approximate the solution of the underlying PDEs. 
Raissi et al. \cite{RAISSI2019686} introduced a physics-informed neural network which takes into account any given laws described by general non-linear PDEs. The authors demonstrated the effectiveness of the proposed framework through application to classical problems in fluids, quantum mechanics and reaction–diffusion systems. Yang et al. \cite{Yangeabd7416} employed conditional generative adversarial neural network to predict stress and strain fields directly from the material microstructure geometry. Wang et al. \cite{wang2021train} introduced a genomic flow network and a mosaic flow predictor to estimate the solution of Laplace and Navier-Stokes equations in domains of unseen shapes and boundary conditions. Pandey et al. \cite{PANDEY20211} proposed a ML based surrogate model for predicting spatially resolved crystal orientation evolution under uniaxial tensile loading. 
It is also important to make sure that the trained network prediction satisfies the fundamental physical and thermodynamics laws \cite{Masi_2021} or fulfills the objectivity and possible material symmetry conditions \cite{Fernandez2021}.

In this scientific context, Convolutional Neural Networks (CNNs) are employed extensively to extract material property-structure relationships based on microstructure images \cite{ZHANG2020113362, LI2019735}. Cecen et al. \cite{CECEN201876} employed a CNN to link a three-dimensional microstructure to its effective (homogenized) properties. The authors showed that the trained CNN is able to learn physically interpretable microstructure features and accurately predict desired properties. 
Rao and Liu \cite{RAO2020109850} proposed a three-dimensional deep CNN to predict the anisotropic effective material properties for representative volume elements with random inclusions. The dataset generated by a computational homogenization approach was used for training the network. They concluded that the trained networks can predict unseen data, indicating that the network is capable of capturing the microstructural features of the system and could produce an accurate prediction of the effective anisotropic stiffness tensor.
Recently, Mianroodi et al. \cite{mianroodi2021teaching} applied a U-Net approach to predict stress fields in geometrically complex and heterogeneous non-linear elasto-plastic material systems. It was shown that the U-Net-based prediction of the stress fields in such a highly non-linear system was about 8000 times faster than that obtained by a typical spectral solver (e.g. \cite{ROTERS2019420}). Interestingly, the U-Net was also capable to reproduce the stress distribution in geometries topologically far from those that had been used for training.

Although the body of literature on multiple possible and promising applications of ML in material science is growing fast, a general approach for transferring constitutive relations directly from atomistic to continuum scale is missing. In the current work, we use Mishin's interatomic potential for Al \cite{Mishin1999} to calculate the elasticity tensor for large sets of randomly generated nanoporous structures. A CNN is then trained using the topology data (stored as images of the atomic structure) and the calculated elasticity tensor components as reference data set. Once the network is trained, it is evaluated in terms of several test cases, showing its ability to capture full atomistic details (such as pore surface effects) while being orders of magnitude faster than the atomistic calculations. The efficiency of the AI-based method coupled with an accurate description of the atomistic scale effects on the material's elastic constitutive response, results in a lossless scale-bridging approach. The preparation of the training data from the atomistic calculations is presented in Section 2. The architecture and training of the CNN is discussed in Section 3. The performance and accuracy of the trained CNN is evaluated in Section 4. Finally, the paper is concluded in Section 5 with a discussion and outlook.

\section{Workflow and Preparation of Atomistic Data}
\label{sec_prepatom}

\begin{figure}[ht]
    \centering
     \includegraphics[width=0.9\textwidth]{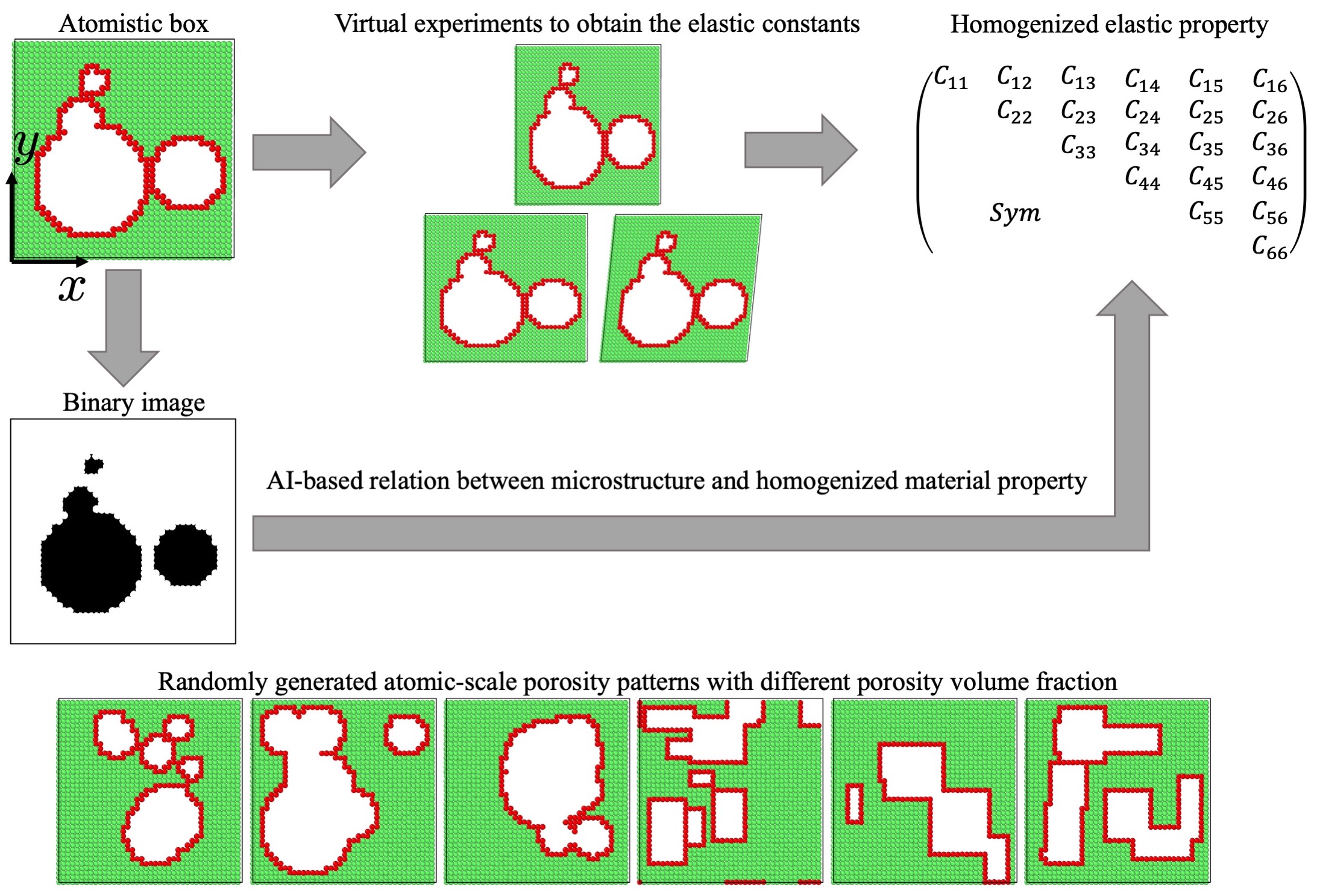}
    \caption{Schematic of the workflow for the elasticity tensor calculation from molecular statics and corresponding image-based artificial intelligence prediction. In this figure and in the others to follow, green and red correspond to the atoms with fcc lattice and unknown (surface) structures, respectively, identified using common neighbor analysis as implemented in the freeware visualization and analysis package Ovito \cite{Ovito}. In the binary image generated from the corresponding atomisitc structure, the pore area is indicated in black while the solid material region in white. Selected examples of the randomly generated circular and rectangular pore structures are shown in the bottom row.
    }
    \label{fig_shem}
\end{figure}

An important step in training a neural network consists in obtaining an accurate and diverse set of training data. In this work we use Molecular Statics (MS) to calculate the anisotropic elastic constants of face centered cubic (fcc) aluminum with randomly distributed porosity. The Mishin embedded atom potential \cite{Mishin1999} in conjunction with LAMMPS \cite{LAMMPS} is employed to calculate the elastic constants. The fully periodic simulation box size is set to $(L_x,L_y,L_z)=(20,20,3)a_0$ where $a_0=4.05$ \AA \, is the lattice constant. The schematics of the method along with exemplary nanoporous structures are shown in Figure \ref{fig_shem}. 

\FloatBarrier
The box is periodic in all directions, representing an infinite array of the pore structure inside the box. Initially the box is filled with $N = 4800 $ Al atoms. Then a random target porosity volume fraction in range of $(0,90)$ percent is selected. The porosity volume fraction is defined as
\begin{equation}
    p = \dfrac{N-n}{N} \times 100,
    \label{eq_porosity}
\end{equation}
where $n$ and $N$ are the number of atoms in the porous and non-porous (completely filled) boxes, respectively. Note that $p$ is 0\% and 100\% in the case of a full simulation box ($n=N$) and empty space ($n=0$), respectively. Once the target porosity volume fraction, for brevity referred to as porosity from here on, is chosen, a number of pores with random shape, size, and position are created by removing atoms from the box until the porosity ($p$) is above or equal the target porosity. The pores are defined as geometrical regions with circular or rectangular shapes with random attributes. To avoid unstable structures, which would complicate the automated MS calculations, all pores are constrained to have minimum distance of 0.4nm from the box boundaries. This will ensure non-zero stiffness in all directions and avoids crumbling structures due to open pores and disconnected parts.

After the desired porosity volume fraction is reached, the atomistic sample is relaxed to zero stress state using a combination of conjugate gradient and FIRE (a damped dynamics method described in \cite{Bitzek2006}) algorithm. Once all stress components (including shear components) are zero, the elastic stiffness calculation is performed using the script in LAMMPS which works by distorting the system infinitesimally in different directions and calculating the change of stress, thereby, measuring the components of the elasticity tensor.

In addition to the porosity volume fraction introduced in \eqref{eq_porosity}, we introduce an atomistic measure for quantifying the pore dispersion
\begin{equation}
    d = \dfrac{n_\mathrm{s}}{N-n} \times 100,
    \label{eq_dispersion}
\end{equation}
where $n_\mathrm{s}$ is the total number of surface atoms, identified using common neighbour analysis \cite{Ovito}. Note that for a given porosity volume fraction $p$, i.e. constant $N-n$, the dispersion $d$ is maximised when the pore surface area, i.e. number of atoms at the surface $n_\mathrm{s}$, is maximised. This typically translates into dispersing the same pore volume fraction into smaller average pore sizes. Note that other quantification measures of porous media such as percolation and dispersion are also available. In the current case, since the structures are randomly generated, at each porosity volume fraction level, we generate many structures with different percolation and dispersion. 

Without loss of generality, we consider first 2D pore structures in the $x\text{-}y$ plane, while along the $z$ direction, no structure variation is assumed, i.e. mimicking an extruded nanostructure. The scheme presented should also work for general 3D nanoporous structures, except that the training of the CNN model becomes computationally more expensive. Thus, only variations in $x\text{-}y$ plane are important here, resulting in the possibility of representing the atomic positions and topology of the pores by a single image. The image is created after full stress relaxation and serves as the input for the CNN. The corresponding outputs are the components of the elasticity tensor. 

Two datasets with circular and rectangular pore shapes are generated, respectively. Each dataset consists of about 10,000 cases, resulting in total 20,000 sets of input image and output elastic constants. Exemplary pore structures selected from each of the datasets are shown in the bottom row of Figure \ref{fig_shem}.

\section{Network Architecture and Training}

Using ML has become ubiquitous in material science (see Refs.~\cite{sha2020artificial, Ye2018,Butler2018,Schmidt2019, Ramprasad2017,Niezgoda2013,McDowell2016,Bereau2016,wodo_broderick_rajan_2016,Agrawal2016,Dimiduk2018, MASI2021104277} for a review). These applications include accelerated material discovery \cite{Raccuglia2016a, Meftahi2020, Sun2019, DAI2020109618}, 
efficient interatomic potential development~\cite{behler2007generalized, Kolb2017,Behler2016,Dragoni2018}, feature identification from complex patterns that have relevance for materials performance  \cite{Jiang2021,Yoshitaka2016,DECOST201730,DECOST2015126,DECOST2017438}, or facilitating predictive simulations which solve macroscopic (non-linear) partial differential equation systems \cite{mianroodi2021teaching,RAISSI2019686,Rad2021a}. 
This has the potential to revolutionize continuum-based simulations of materials, allowing a substantial enhancement in the modeling of systems and topologies with high complexity.
Macroscopic simulations (with or without AI) require homogenized and averaged  material property descriptors, e.g. elastic constants and yield strength, as their inputs. In a multi-scale approach these macroscopic properties are estimated using microscopic simulations like molecular dynamics or \textit{ab initio} calculations \cite{Fernandez2020}. Interestingly, ML has been also applied to replace  (at least partially) the expensive microscopic simulations. These efforts include replacing the computationally costly interatomic \textit{ab initio} force calculations by a neural network \cite{Parrinello2007}, or replacing the trajectory prediction using graph neural networks \cite{deepmind2020}. A common feature of these ML solutions is that the atomistic coordinates are transferred to the network as its input. 

Unlike these approaches, in our case we use the image of the microstructure as the input and predict the aggregated effect of the interatomic forces and the arrangements of the atoms. In this respect, our work is similar to the approach  introduced in \cite{Naif2020} where 2D X-ray images of the microstructure are used as the input and macroscopic material properties like porosity, specific surface area, and average pore size of each image are computed. CNN models of homogenized mechanical properties have been trained in relation to mesoscopic microstructure with multiple phases \cite{ZHANG2020113362, LI2019735,CECEN201876}, whereby the classical continuum theory was applied to each phase and for the evaluation of the effective properties. Our approach differs fundamentally from these CNN models in the sense that we apply atomistic calculations to obtain the property which automatically map the surface effects, namely, the elastic stiffness changes resulting from the surface reconstructions of the many atoms surrounding the pores, and thus can recapitulate the resultant size dependency. 

\subsection{Neural network architecture}
\label{sec_nn}
\begin{figure}[!ht]
    \centering
     \includegraphics[width=0.8\textwidth]{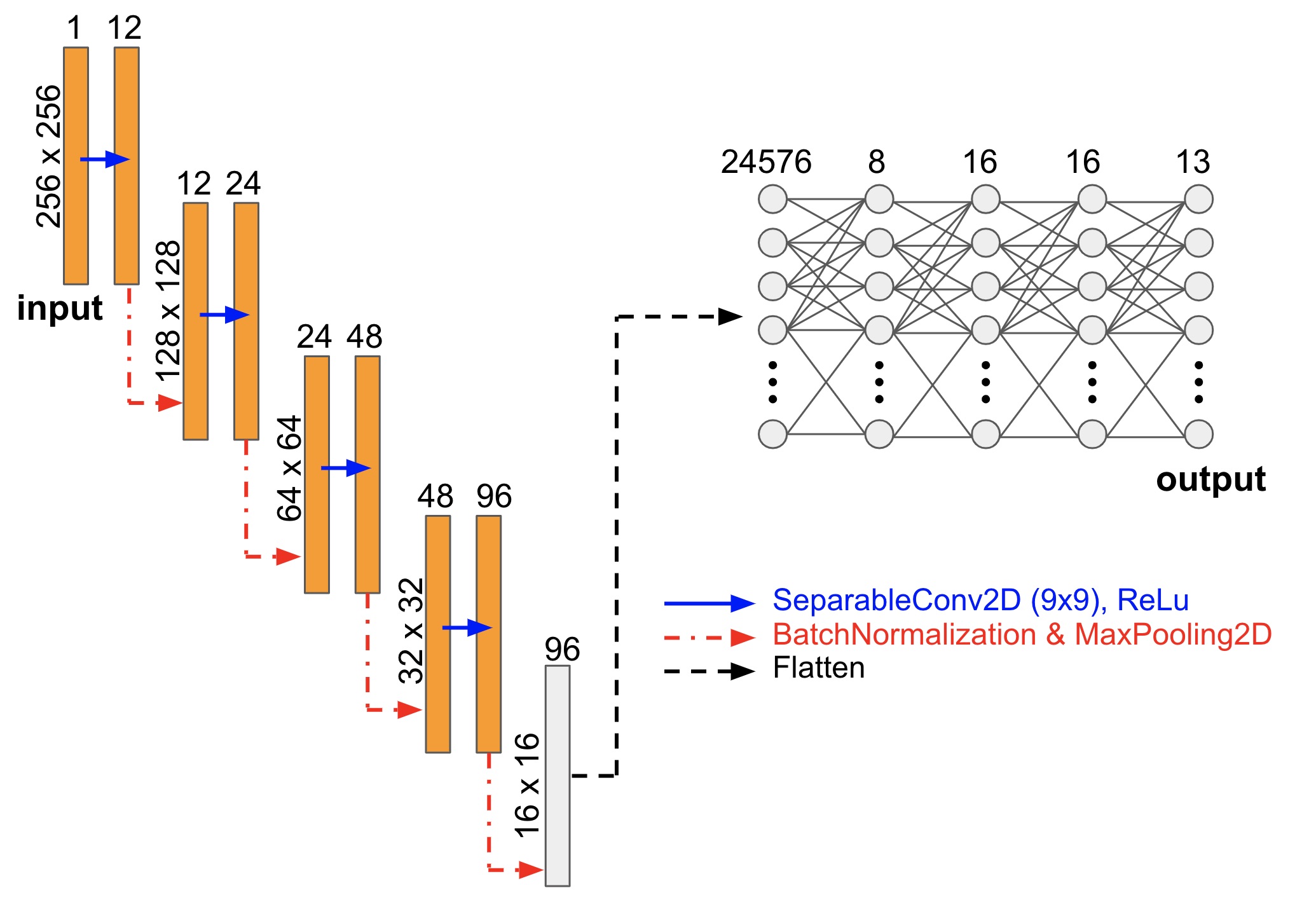}
    \caption{The artificial neural network architecture used in this work consists of a convolutional part (left) and dense layers (right). The input is a 256$\times$256 pixel binary image and the output is a set of 13 scalar values of the elasticity tensor.}
    \label{fig_network}
\end{figure}

Neural networks vary in their basic neural units, the arrangement of these units in the layers and their connectivity, the character of the loss functions (e.g in terms of the quantification of the deviations of the predicted values relative to the reference values in the training data), and the 'reductionist' spirit of the network design (e.g. see Refs.~\cite{hochreiter1997long, ronneberger2015u, lechner2020neural}). 
We build our neural network under the assumption that the macroscopic properties can be obtained as a non-linear function of the coarse-grained local information. In other words, the network extracts the local information, coarse grains it, and finally passes this information to a (learnable) non-linear function. This is reflected in the architecture of our neural network, as shown in Figure \ref{fig_network}.  

Given that (i) the input to our network is an image and (ii) the first step is to extract the local information, we use convolutional layers which are the most widely used neural network architecture for image processing \cite{chollet2018deep}. We apply, successively, a number of convolutional layers where each layer is accompanied by a coarse-graining step which is implemented by a max pooling operation. The convolutional layer is commonly coded in the form of a simple matrix (of much lower rank compared to the image size). This matrix, which is commonly referred to as the kernel, is sequentially slid across the image and multiplied at each sequential position with a subset of the input array such that the output enhances certain topological pattern features such as edges, corners, gradients, etc. As a rule of thumb, including more kernels in the design of the network leads to a better performance of the network; simply put, the network then has a larger potential for detecting features. As mentioned before, candidate quantities for these features could be gradients, edges, etc. Note that these features are not hard-coded. Instead, the network learns by itself what the important features are. The process of learning the elements of these kernels is as follows. The elements of this matrix are initially set to be random numbers but during the training phase, the neural network learns an appropriate set of values for these elements such that (together with the rest of the network) the prediction error is minimized. To enable the network to obtain different types of (necessary) information, a number of kernels are used. After feature extraction and coarse-graining, the most coarse-grained information is mapped to the homogenized properties via a number of dense layers, which is essentially a non-linear transfer function. 

The neural network that we utilized here is very similar to the \textit{encoder} part of the U-Net architecture \cite{ronneberger2015u}, which has shown to be a good candidate for capturing solutions to solid mechanics PDEs~\cite{mianroodi2021teaching}. Generally speaking, the encoders map the information from one space to another, where the second space is smaller in its dimensions. Later, this condensed information is passed to the decoder whose task is to map the information from the low dimensional space back to a high dimensional space. In the case of the U-Net, the high-dimensional input is an image (which is for example a RGB image showing an object). This image is transferred to the encoder and condensed to a much smaller image. Then, the condensed data is sent to the decoder which creates an image with the size of the original image (which is for example a mask indicating the positions of that object in the original image). In the current approach, we replace the decoder by a non-linear function which maps the encoded representation to the scalar output values (the elasticity tensor components) which we are interested in.

\subsection{Neural network training}
In the current implementation, for the convolutional layers a kernel of size  $k\times k$ with $k=20$ is used. The kernel size used here is considerably larger (i.e. $\simeq$ 6 times) than that used in the original CNN design. The rationale behind using a larger size is to capture more non-local effects. The number of channels progressively increases from one channel input data to the embedded space which has 96 channels as shown in Figure \ref{fig_network}. After that, we use five dense layers in conjunction with a Rectified Linear Unit (ReLU) activation function which are accompanied by the last dense layer with no activation function (i.e. the identity activation function). The training is performed by minimizing the loss function which measures the mean squared difference between the predicted and atomistically calculated values across all the 13 outputs. Note that, as it will be explained in the next section, only 13 components of the full elasticity tensor are independent in the current case. This error minimization is done using the ADAM optimizer  \cite{kingma2014adam} which is a stochastic first-order gradient-based method. The parameters of the ADAM optimizer are set to $\beta_1=0.9$, $\beta_2=0.999$, $\epsilon=10^{-7}$, and a learning rate within the range $[1,2] \times 10^{-4}$. We use random samples in batches of size 128 for the gradient estimation, and continue training for 1000 epochs. In each epoch, the complete training dataset is used. The data is divided into training and validation sets as explained in the next section. 

\section{Results}

\subsection{Atomistic results}
\label{sec_AtomRes}

\begin{figure}[ht]
    \centering
     \includegraphics[width=0.99\textwidth]{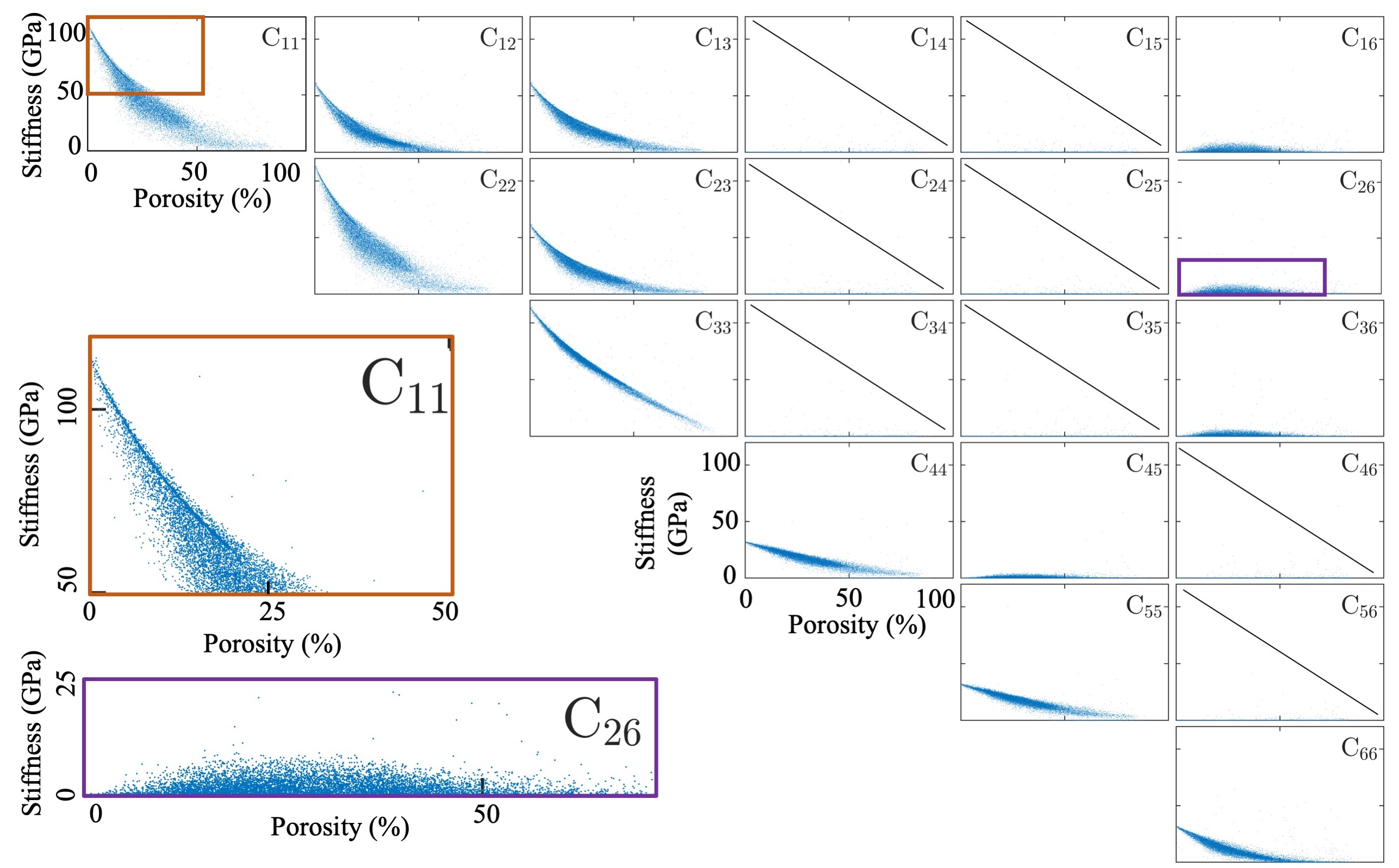}
    \caption{Elasticity tensor components for 18,172 atomistic boxes with randomly generated pore structure as a function of the box porosity volume fraction, defined in \eqref{eq_porosity}. All the plots (except the zoomed versions of $\mathrm{C}_{11}$ and $\mathrm{C}_{26}$ at the bottom left) have the same range in horizontal (porosity) and vertical (stiffness) axes. }
    \label{fig_Data}
\end{figure}

From the 20,000 randomly generated structures and the resultant elasticity tensors, only those with positive definite elasticity tensors are selected. Since the pore structures are random, in some cases, the geometry is not stable, leading to incorrect elasticity calculations. After filtering these unacceptable cases, 18,172 samples with physically meaningful input and output data are used for training the CNN. All of these data is visualized in Figure \ref{fig_Data}. Note that the randomly shuffled dataset is divided into training and validation (test) subsets with sizes of 17,172 and 1000, respectively. 

As observed in Figure \ref{fig_Data}, the dispersion of the data is increasing as the porosity is increased. This is expected as there are more random ways to create the same porosity volume fraction when it it increased. 
Also note that more of the cases in high porosity regions are filtered out due to the instability of the generated structures. Therefore, the number of data for structure regimes with higher porosity is lower compared to those with low porosity.
As explained in Section \ref{sec_prepatom}, randomly generated structures with higher porosity fractions are often unstable and cause issues for the automated elasticity tensor calculations. Therefore, there is an unbalanced distribution of data at different porosity levels. As it will be shown later, this results in a reduced prediction accuracy for the higher porosity levels with lower number of training data. 

As it is shown in Figure \ref{fig_Data}, only 13 components are non-zero (within tolerance of $10^{-3}$GPa), instead of 21 for a general elasticity tensor. This is due to the assumed 2D variation of the pore structures and translates to the introduction of a symmetry-breaking effect, rendering the structure response monoclinic (rather than cubic) (with a plane of symmetry) \cite{SADD2021331}. Note that the Voigt notation, i.e. $(11,22,33,23,13,12) \rightarrow (1,2,3,4,5,6)$ is employed. Therefore the CNN has 13 scalar outputs for one input image as summarized above. The magnified views of the $\mathrm{C_{11}}$ and $\mathrm{C_{26}}$ plots are also shown in Figure \ref{fig_Data}. As apparent for the $\mathrm{C_{11}}$ plot, there is a high density of points along a curve from 0\% porosity up to about 20\%. This high point density portion of the curve is visible as a line shadow in this region which seemingly disappears for porosity values above 20\%. This can be due to the percolation threshold, i.e. as the porosity volume fraction increases, at a certain threshold (which highly depends on the structure) most of the pores will be connected to each other increasing the dispersion in the data.

For the $\mathrm{C_{26}}$ elastic constant, the maximum dispersion of the data is around 20-25\% porosity while the data points are closer to each other for porosity values around 0\% and above 50\%. A similar trend is visible for other components such as $\mathrm{C_{16}}$ and $\mathrm{C_{36}}$. This could be also related to the percolation threshold discussed above. 

\subsection{Benchmarking AI predictions}

The overall mean square error of the trained network on the test dataset is about 7.6GPa, which corresponds to about 9.6\% of the equivalent bulk modulus of single crystalline Al. However, besides the 1000 cases of the atomistic data reserved for the test dataset as explained above, we also look into a few special cases here to evaluate the performance of the network. Note that these cases were neither part of the training nor the test dataset.  

\subsubsection{Size effects}
\label{sec_sizeeffect}
\begin{figure}[ht]
    \centering
     \includegraphics[width=0.65\textwidth]{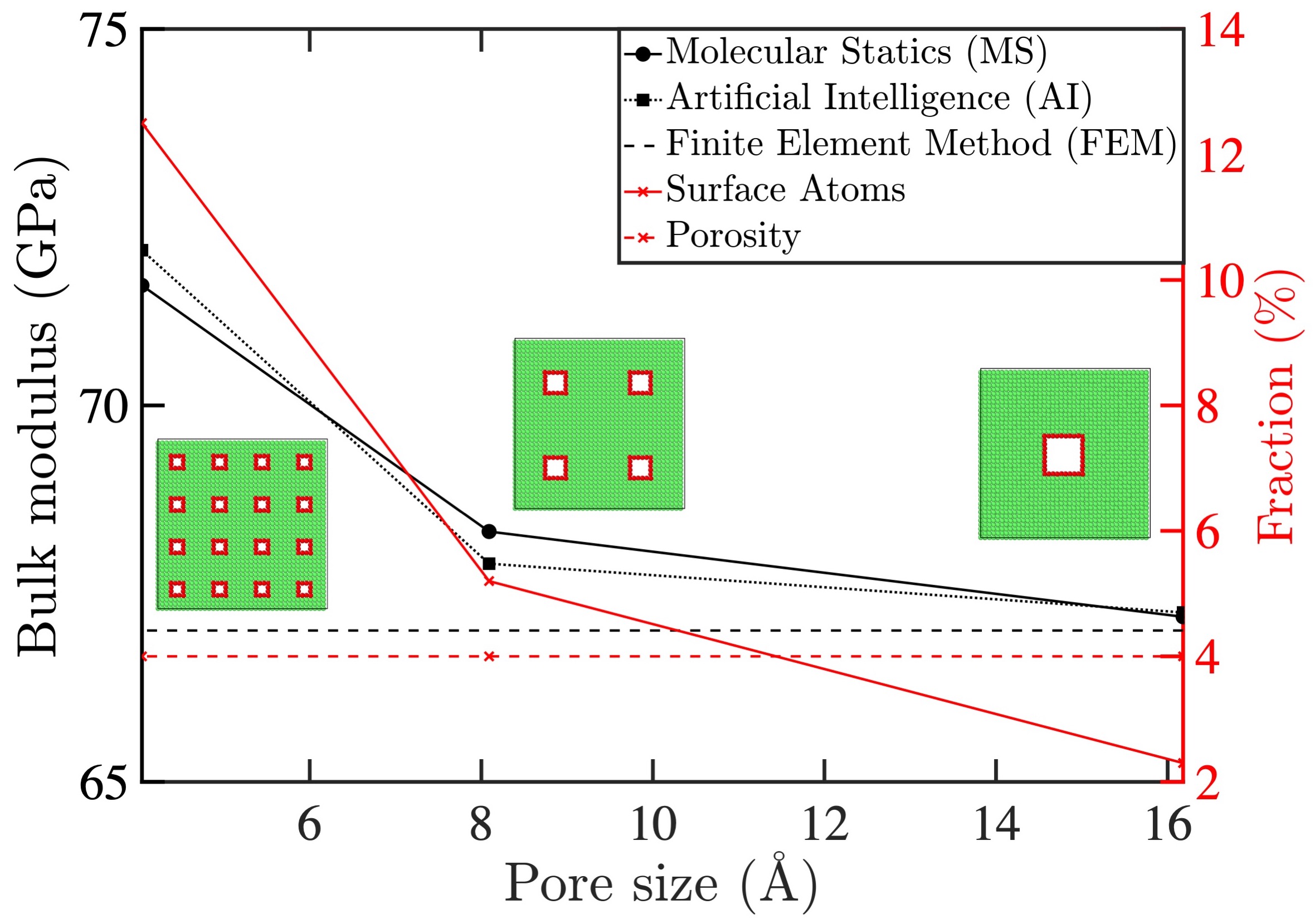}
    \caption{Effective bulk modulus (black) as a function of pore dispersion in three simulation boxes shown. Classical linear elastic continuum mechanics (solved by FEM, dashed black) shows no size effect, while atomistic calculations (solid black) and AI predictions (dotted black) show a significant size effect. Note that the pore volume fraction (dashed red) is constant for all three cases while the fraction of atoms that are located at the inner surfaces of the pores (solid red) increases as the pores get smaller, an effect which leads to a stiffer response.
    }
    \label{fig_Sizeeffect}
\end{figure}

One major difference between continuum mechanics and atomic systems is the non-local nature of the interactions in the latter. Due to the quantum-mechanical nature of their electronic bonding, atoms (particularly in metals) interact over a wide distance (where in simulations a cutoff radius is usually defined) as opposed to local material models used in classical solid mechanics. This non-local interaction leads to the distinction between atoms at a free surface with broken neighbor bonds and atoms in the perfect bulk fcc lattice where all bonds are saturated. As systems become smaller, the ratio of atoms at surfaces to the atoms in the bulk increases (see red solid line in Figure \ref{fig_Sizeeffect}). To demonstrate the performance of the trained CNN, we investigate the size effect dependency of the effective bulk modulus of a system with a pore and compare the results from the MS simulations, as well as classical continuum mechanics solved using the FEM in conjunction with a linear elastic constitutive law, without any phenomenological size effect considered. Note that the bulk modulus in this work is the equivalent bulk modulus of single crystalline Al with arbitrary pore structure. 

To this end, three boxes with square shaped pores are considered as shown in Figure \ref{fig_Sizeeffect}. In all cases the simulation size and orientation is the same as explained in Section \ref{sec_prepatom}. The pore in simulation box on the right hand side of the figure has the edge length of $L_x/5$, resulting in a porosity volume fraction of  $p=4\%$. In all other steps, the edge length of the pore is divided by two (area reduced by a factor of four) while the number of pores are quadrupled. Therefore, $N-n$ and the porosity volume fraction, see \eqref{eq_porosity}, of the three boxes remains unchanged (dashed red line in the figure) while dispersion parameter $d$, see \eqref{eq_dispersion}, is increased from 56.3\% to 125.0\% and 300.0\% in the three cases, as the pore sizes become smaller. 

The same input structure is fed to the trained CNN and the predicted bulk modulus is plotted in Figure \ref{fig_Sizeeffect} as solid black squares. As observed, the CNN prediction captures the pore size dependency of the bulk modulus with good agreement compared to the MS calculations. Note that the trained CNN is predicting the 13 components of the elasticity tensor as discussed above. The effective bulk modulus is calculated using
\begin{equation}
    K = - V\dfrac {dP}{dV} \approx - V\dfrac {\Delta P}{\Delta V},
      \label{eq_bulk_def}
\end{equation}
where $V$, $\Delta P = (S_{11} + S_{22} + S_{33})/3$ and $\Delta V=V(1+E_{11})(1+E_{22})(1+E_{33})-V$ are the volume, and the differential of the hydrostatic pressure and volume, respectively. Stress $\bm{S}$ and strain $\bm{E}$ are related through the elastic relation $\bm{S} = \msbi{C} \bm{E}$. Under the small deformation assumption and with initially undeformed / stress-free conditions the differentials can be approximated as
\begin{equation}
    \Delta V  = 3\epsilon V, \,\, \Delta P = (S_{11} + S_{22} + S_{33})/3,
    \label{eq_diffs}
\end{equation}
where $E_{11} = E_{22} = E_{33} = \epsilon$ is assumed. Combining \eqref{eq_bulk_def} and \eqref{eq_diffs}, we get
\begin{equation}
   K \approx -\dfrac{1}{9\epsilon} (S_{11} + S_{22} + S_{33}).
   \label{eq_bulk}
\end{equation}
The bulk modulus is then obtained employing  \eqref{eq_bulk} and applying an equal triaxial strain state on the system. Such displacement boundary conditions were applied for both, the  MS and the linear elastic FEM calculations. In the FEM case, the solid phase is equipped with the anisotropic elastic property of single crystalline Al while the pore has zero stiffness. Note that in all the simulations and results reported in this paper, we work with single crystalline material and a full anisotropic elasticity model. Homogenized polycrystalline properties could be extracted from these for any given grain structure \cite{ROTERS20101152}. 

The classical continuum prediction for the single crystal bulk elastic modulus of this system is size independent (dashed black line in Figure \ref{fig_Sizeeffect}) with a value of 67.0GPa. However, the MS calculations show a size effect in the bulk modulus with an increase from 67.2 to 71.6GPa (6.5\% increase) as the pores get smaller from 16.1 to 4.0\, \AA\, (solid black line in the figure). This is correctly captured by the trained CNN with a relative error of about 0.6\% compared to the MS-based values. 

\subsubsection{Porosity and connectivity}

\begin{figure}[ht]
    \centering
     \includegraphics[width=0.99\textwidth]{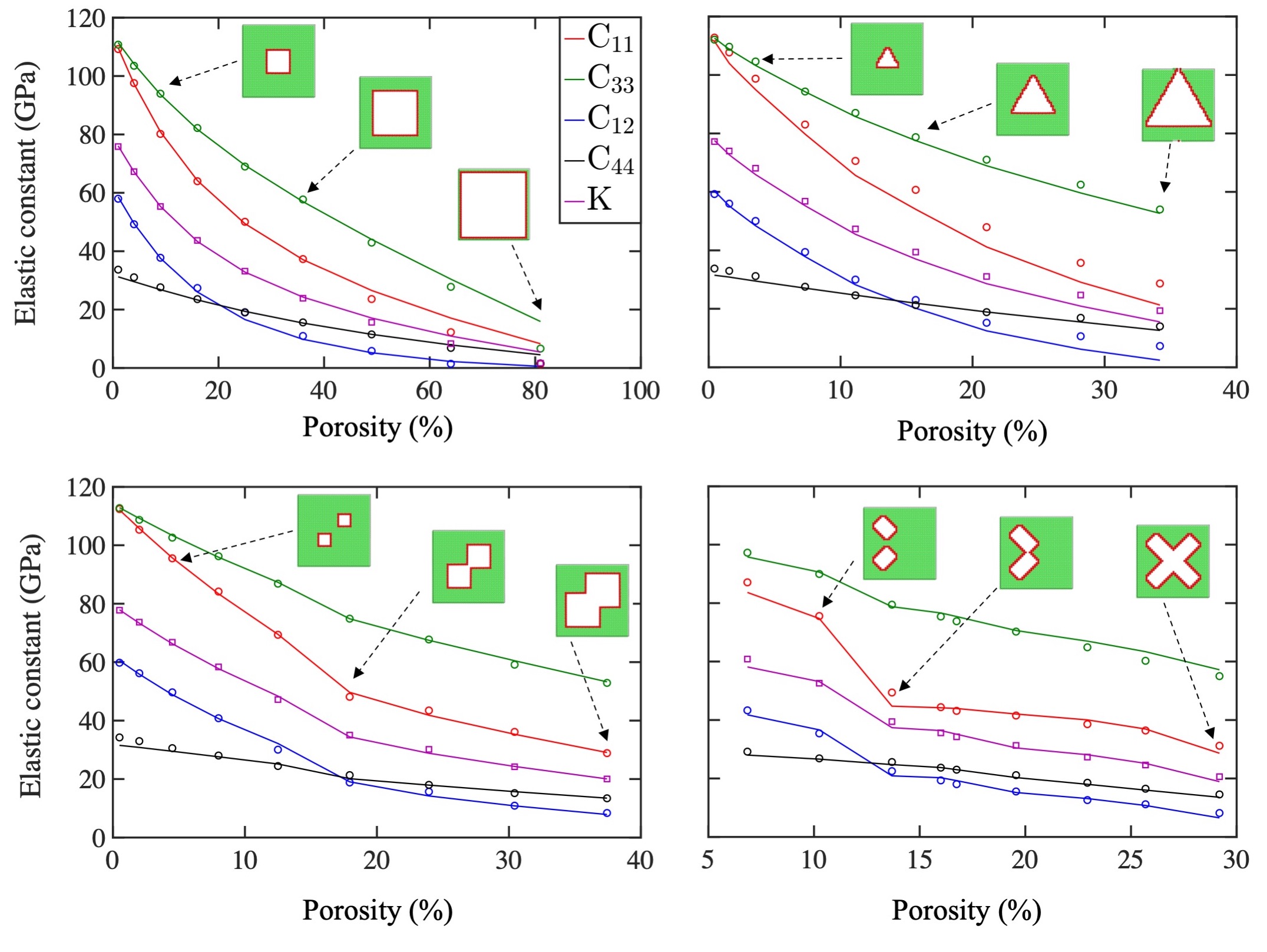}
    \caption{Atomistic calculations (solid lines) and AI predictions (markers) for selected components of elasticity tensor and effective single crystal bulk modulus for cases with simple geometries of one square pore (top left), triangular pore (top right), two square pores (bottom left) and two rectangular pores (bottom right).
}
    \label{fig_3cases}
\end{figure}

Here, we systematically evaluate the CNN-based prediction of various components of the single crystal elasticity tensor against the MS results for special cases shown in Figure \ref{fig_3cases}.

As seen in Figure \ref{fig_3cases} top left for the case of a square-shaped pore, the agreement between MS (solid lines) and CNN prediction (marked points) is excellent. For example, considering the $\mathrm{C_{33}}$ results, the relative error starts from 0.6\% for the smallest pore size and increases to 8.7\% for the second largest pore (before the last one). The CNN prediction deviates about 58\% from atomistic calculations in the last point of the comparison at a very large porosity of 81\%. The deviation is attributed mainly to the reduced number of training data provided at such high porosity levels as discussed in Section \ref{sec_AtomRes}. Furthermore, as explained above, the training data exclude all pores that were positioned closer than 4\AA\, to the boundary of the box, which explains the larger deviation between the AI prediction and MS results at the highest porosity volume fraction in this case. 
The CNN prediction also shows a relatively good agreement with the MS results in the system with a triangular pore as shown in the right top panel of Figure \ref{fig_3cases}. Although the error of 1.1\% for the initial microstructure is increased to 34\% for the largest pore, the network still captures the important trend in the results. 
The higher error compared to square-shaped pores in the previous case is probably due to the fact that no triangular or similar shaped pores were included in the randomly generated training dataset, as opposed to random circular and rectangular shaped pores.  
In the case of two growing squares in the bottom left panel of this figure, an interesting change in the slope of the plot (in particular for $\mathrm{C_{11}}$) is visible. This change in the slope is due to the two squares reaching each other and overlapping after this point. The change in slope is correctly captured by the CNN prediction signifying the effectiveness of the trained network for including effects of topology, particularly of pore percolation. Since the maximum porosity in this case is up to 37\%, the CNN prediction agrees much better (compared to previous cases) with the MS results with a maximum relative error of 2.2\%. 
A similar test for two rectangular pores of increasing size has been performed with the results shown at the bottom right panel of Figure \ref{fig_3cases}. Similar to the previous case, the overlapping of the two rectangles leads to changes in the slope of the plots. The agreement between CNN predictions and MS results is quite good with maximum error of about 5.0\%. The changes in the slope of the plots indicate the effect of connectivity of the pores on the data, which seems to be captured by the CNN correctly. 

\begin{figure}[ht]
    \centering
     \includegraphics[width=0.95\textwidth]{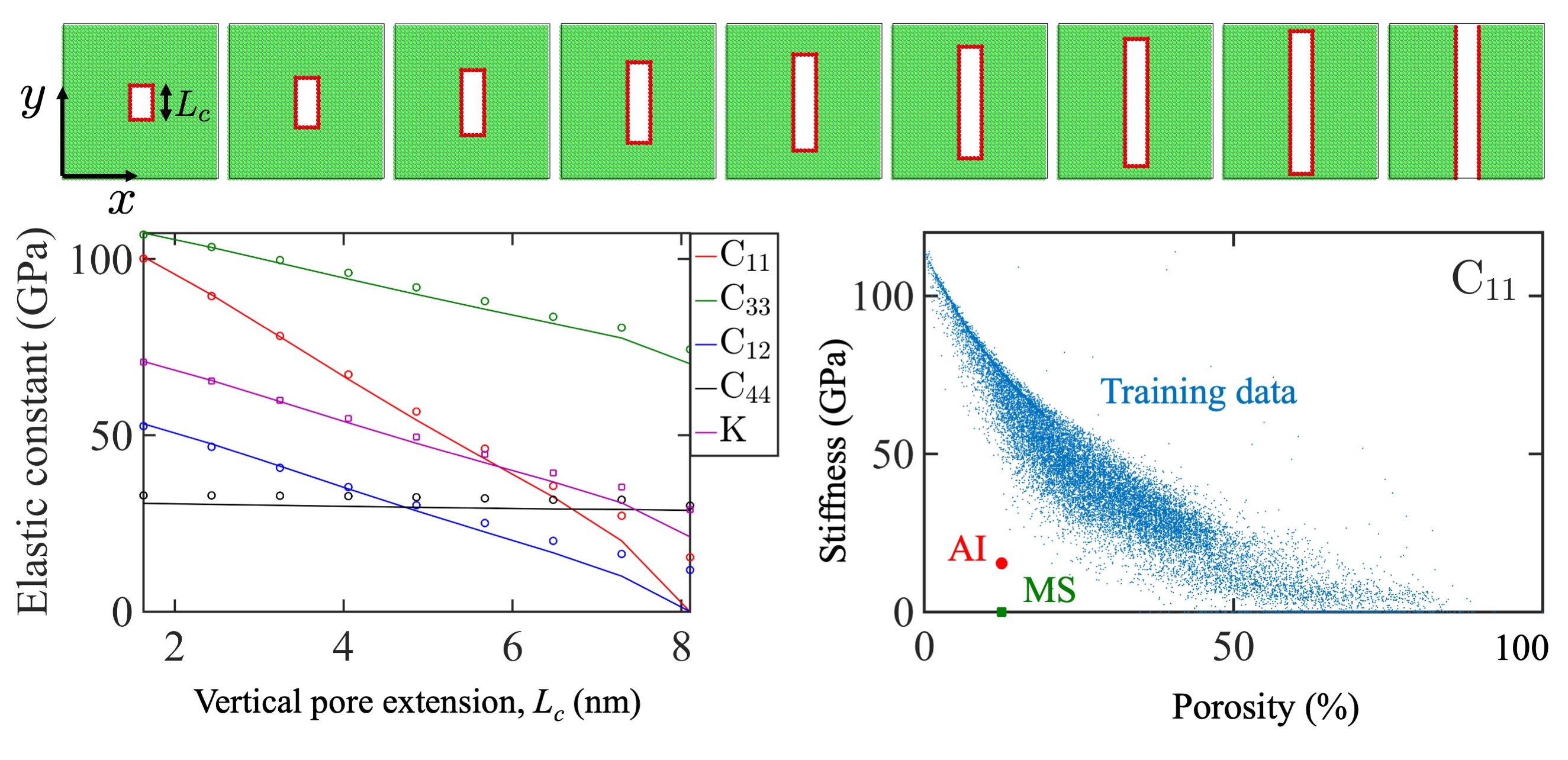}
    \caption{Top row: Simulation boxes with vertically growing rectangular pore of length $L_c$.
    Bottom-left: MS calculation (solid lines) and AI prediction (markers) for single crystal elasticity tensor components in a vertically growing rectangular pore as a function of $L_c$. Bottom-right: The $\mathrm{C}_{11}$ training data as well as the MS calculation and AI prediction for the case of $L_c=8.0$nm (bottom right). }
    \label{fig_extrapolate}
\end{figure}

Next, we study a case of a vertically growing pore with both, CNN and MS calculations. As shown in Figure \ref{fig_extrapolate}, this case is analogous to a crack growth, where depending on the crack orientation, a strong asymmetry for the elasticity components is expected. Some components of the elasticity tensor drastically decrease (e.g. $\mathrm{C}_{11}$) while others remain almost constant (e.g. $\mathrm{C}_{44}$) or decrease with a lower rate (e.g. $\mathrm{C}_{33}$), as the vertical pore length increases. 

According to Figure \ref{fig_extrapolate} bottom left, the agreement between MS and CNN-based prediction is apparent for smaller pore lengths. 
The trained model is also sensitive to the pore orientation and captures the stiffness degradation correctly. As the pore becomes longer in the vertical direction and therefore gets closer to the box boundaries, the error of the CNN prediction increases. In the limit of a fully separated box ($L_c = 8.0$nm), as expected, the MS prediction for $\mathrm{C}_{11}$ provides a complete loss in stiffness, whereas the AI prediction for this case is $\mathrm{C_{11}}=15.4$GPa. This case, as shown in Figure \ref{fig_extrapolate} at the bottom right, is completely out of the range of the training data. However, the AI prediction, although associated with high error, is also out of the training data range and qualitatively points in the correct direction. For better evaluation of fully grown pores (which could for instance mimic cracks or delamination features) and separated boxes, the training data should include such cases. However, in this work, we only focus on cases with non-zero stiffness and non-fractured boxes. 

\subsection{Lossless multi-scale modeling}

The CNN prediction of the elasticity tensor has been incorporated in an FEM model as a continuum scale constitutive law to simulate bending of a nanoporous beam. The beam dimensions are $(l_x,l_y,l_z) = (405.0, 81.0, 1.2)$nm with free surfaces in $x$ and $y$ and periodic boundary condition along the $z$ axis. The left side of the beam is fixed in all directions while the right side is displaced step-wise in $-y$ direction up to $|u_y| = 40 \mathrm{nm}$. The atomistically modeled beam has square shaped pores with side length and spacing of 0.40 and 1.46nm, respectively, and consists of 2,304,480 atoms. Note that the porosity of the beam corresponds to the smallest pore size studied in Section \ref{sec_sizeeffect}.

\begin{figure}[ht]
    \centering
     \includegraphics[width=0.95\textwidth]{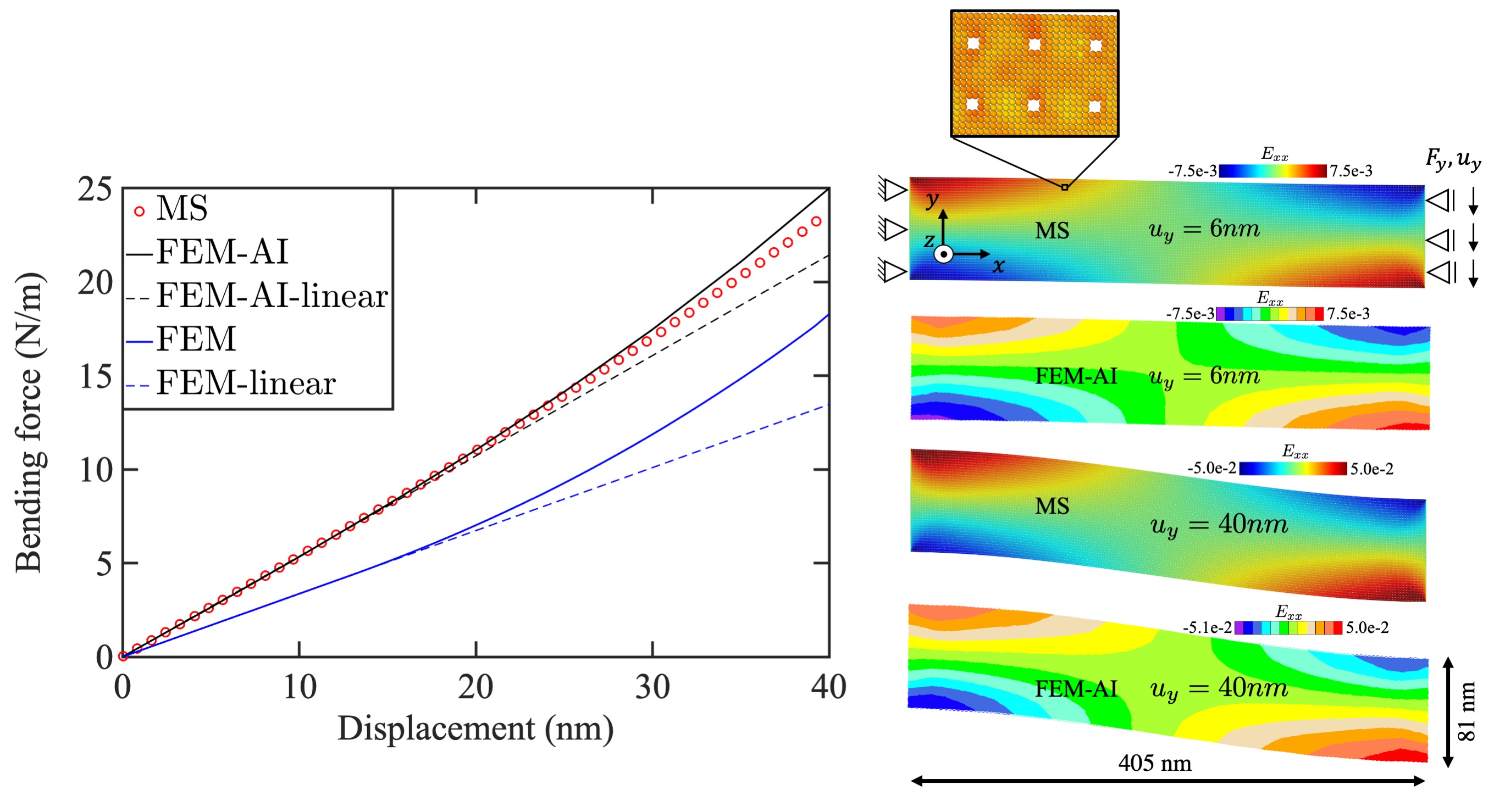}
    \caption{Comparison of force per unit length in $z$ of the beam as a function of displacement in $y$ (left) based on fully atomistic simulations (red markers) and a large deformation finite element calculation equipped with the AI-based constitutive elastic relation (black line), as well as the FEM model with size independent constitutive relation (blue line).  
    The dashed lines are the system response ignoring the large deformation effects.
    Strain fields calculated at two different beam deflections of 6nm and 40nm based on MS and FEM-AI are shown on the right side. }
    \label{fig_bend}
\end{figure}

The same beam geometry is simulated by FEM in conjunction with a full anisotropic elasticity tensor predicted by AI for this porosity. The beam is then loaded in a geometrically non-linear FEM simulation up to $|u_y|=40nm$. The resulting force-displacement curve agrees well with the results obtained from the full atomistic simulations. Note that, as shown in Section \ref{sec_sizeeffect}, this pore size introduces a significant surface/size effect compared to a continuum pore. As seen from the results, using the AI-based constitutive model, seems to correctly transfer the mechanical behaviour of the system from the atomistic scale into a continuum model.

The results of the FEM simulations obtained by using Hooke's law for the linear elastic constitutive response, i.e. without the AI-based size-dependent elastic model, are shown in Figure \ref{fig_bend} left as blue curve, for the same pore distribution. As it is seen here, ignoring size dependency of the elastic stiffness introduces a significant error in the predicted bending behaviour of the beam. For example, at $|u_y|=20$ nm, MS and FEM-AI both predict a bending force of 11.0 N/m while the size-independent FEM calculation predicts a bending force of 7.0 N/m.
Note that the surface effects increase the stiffness of the system as shown in Figure \ref{fig_Sizeeffect}. Therefore, as expected, the atomistic simulation and the FEM simulation in conjunction with the AI-based elastic constitutive model predict higher forces for the same beam displacement compared to the FEM with no surface effects.
The deviation (about 5.2\% at $|u_y|=40$ nm) between the MS and the FEM-AI simulations for larger displacements is partly due to the increased pore deformation and its effect on the elasticity tensor. The FEM-AI case at the moment does not update the pore shape to reevaluate the elasticity tensor during the loading. However, this will be straightforward to implement and expected to resolve the deviation at larger deformation values as well. 

For a better comparison and to emphasize the potential of the AI-based multi-scaling constitutive modeling capability, we compared the computational time required  for different approaches as well as different steps of the training in Table~\ref{tab:timecomp}. In the first row, the required computational times for training the network (as explained in section \ref{sec_nn} with the data in section \ref{sec_prepatom}) is listed. Next, the computation time for one elasticity tensor calculation with MS and AI is compared. The bottom two rows of the table list the computation times required for the nanoporous beam bending simulations as discussed above. 

\begin{table}[H]
	\centering
	\begin{tabular}{ l l l } 
                              & Time (sec)   & System    \\ \hline
	AI training               &$2.0\times 10^5$ ($\sim$2 days) & NVIDIA Tesla P4 GPU \\ \hline
    \multicolumn{3}{l}{Elasticity tensor calculation}  \\ 
	MS &$101$      & Intel i7-860  2.80 GHz (single core)\\
	AI   &$0.43$    & Intel i7-860  2.80 GHz (single core)\\
	\hline          
    \multicolumn{3}{l}{Nanoporous beam bending test}  \\ 
    MS  &$2.03\times 10^6$ ($\sim$23 days) & Intel Xeon 6150 2.70 GHz ($\times 2$, 36 cores) \\
    FEM-AI          &$35.5$   & 1 CPU (averaged over 10 simulations)  \\
	\hline

	\end{tabular}
	\caption{Computational time for training the network and a single elasticity tensor calculation with MS and AI, as well as the beam bending test (bottom two rows) with different approaches. }
	\label{tab:timecomp}
\end{table}

As listed in Table~\ref{tab:timecomp}, although the training time of the CNN is about two days, the evaluation of the trained network for the prediction of the full elasticity tensor is about 230 times faster than the MS calculations. This significant speed-up in elasticity tensor calculation will benefit FEM-MS-based multi-scale modeling, where each integration point of the elements in the continuum scale FEM model is represented by an atomistic simulation box. Since in the current simple example of a uniform pore structure in the beam exposed to a bending load, an on-the-fly evaluation of the AI-based elasticity tensor was not necessary, the FEM method with the AI-calculated constitutive relation (FEM-AI) is significantly  (about five orders of magnitude in this particular example) faster
than the full atomistic (MS) calculation, while still capturing important size effects as shown in Figure \ref{fig_bend}. Note that the speed-up in this case highly depends on the problem complexity and the numbers reported in the lower part of the Table~\ref{tab:timecomp} are only representative of the current case and presented here as a first point of reference. The main speed-up reported is the calculation of the fully anisotropic elasticity tensor with AI as compared to the time-consuming MS methods.
It should be also considered in this context that the elastic response is the most simple, fast and straightforward mechanical response feature to predict in such atomistic calculations.  When addressing more complex size-dependent nanostructural features, such as for instance inelastic response of materials under load,  atomistic calculations are computationally much more costly,  so that corresponding AI approximations of the size dependence (or other types of) underlying and homogenized constitutive response might profit even much more from such methods.

\section{Conclusions and Outlook}
\label{sec_conclusion}

In this work, we show a lossless and efficient approach to use AI techniques to derive constitutive laws for multi-scale modeling of materials with complex nanostructures. Two ideas are presented, namely, (1) using computations at the atomistic scale and transfer the information to the continuum level to arrive at physics-motivated model parameters without any ad-hoc assumptions or empirical approximations. (2) A Convolutional Neural Network (CNN) is used to directly relate the image of an arbitrary complicated material structure to its homogenized properties. More specifically, we focused on obtaining the anisotropic elastic properties of a nanoporous aluminum single crystal, as a reference model substance, which can be well simulated by atomistic methods. Random porous structures are generated and their homogenized anisotropic elasticity tensor is calculated using Molecular Statics (MS). We show that all the non-local effects arising from the physical nature of the material's response at the atomistic scale (i.e. surface and size effects) are captured in this step. The complex pore structure and the details of the atomistic reconstructions at the pore surfaces, necessitates a neural network design that can detect the local properties and aggregate that information for use at a coarse-grained scale. The convolutional neural networks are one of the few appropriate candidates to achieve this flow of information. The trained CNN confirms this by adequately predicting the elasticity tensor of completely new porous structures which were not included in the training set. Finally, the proposed methodology is applied to a simple structural problem of a beam under bending load. We show that by using a Finite Element Method (FEM) simulation in conjunction with the  AI-based constitutive relations, one can efficiently and accurately predict the material's structural behavior (in the elastic regime, with full size dependence) with the accuracy of full atomistic simulations. We also observed that the trained AI can predict anisotropic elastic properties about $230$ times faster compared to performing corresponding MS calculations. Based on the latter observation, one can also conclude that such materials-related accelerated and lossless AI-based multi-scale modeling will have at least the same amount of speed-up compared to a conventional hierarchical multi-scale simulation combining FEM and MS. 

Although the performance of the proposed approach using the investigated dataset is significant, there is room for further improvements. As an example, it would be interesting to enrich the current training data set by including more cases with much higher porosity, including zero stiffness cases as well as open-cell topologies. Finally, instead of training the symmetric part of the stiffness tensor, one can try to predict the Cholesky factor of a tangent stiffness matrix to impose a weak convexity on the strain energy function \cite{XU2021110072}. The latter point is specifically interesting when it comes to non-linear material behavior.

The present work opens up many opportunities for efficient and yet accurate inverse design strategies.
For future work, the current methodology can be extended to study similar engineering materials with complex inherent porosity features such as metamaterials, biological matter, additively manufactured materials, batteries or foams \cite{DESHPANDE20011747, Kulagin2020, LIU2020104060}. Investigations should be done to compare the introduced AI-based model with more advanced non-local elasticity models such as micropolar \cite{Rueger2016} or peridynamics \cite{GUO202148} theory. Furthermore, instead of finding the symmetric part of the stiffness tensor as an output, one can focus on obtaining the total scalar elastic energy as a function of the given deformation gradient. In this manner, one can also derive physics-based material non-linearities for hyperelastic response.
The proposed approach may also be extended to perform studies on arbitrary (nano-) composite volume elements with various phases. As a result, one can obtain not only the mechanical properties (e.g stiffness tensor) but also other physical properties such as effective thermal conductivity or the mobility of chemical species in such complex types of matter. The reported collection of data can also help to understand how the porosity with an arbitrary shape influences the overall stiffness. Such information could be further used for developing damage models at coarse-grained scales. In other words, the structure dependent change in the stiffness tensor, for instance due to creep porosity or micro-fracture patterns, can be used to represent the level of damage or degradation due to the change in porosity within each material point.
Finally, other fundamental and essential properties such as direction-dependent fracture energy \cite{REZAEI2021a} may be extracted from atomistic simulations. The anisotropic fracture energy together with anisotropic elastic properties can be directly passed to macroscopic phase-field damage models at larger scales for conducting efficient calculations based on atomistic data.

\section*{Acknowledgements}
We thank Prof. Bob Svendsen for fruitful discussions during this work. 

\bibliographystyle{naturemag}

\bibliography{Ref.bib}

\end{document}